\documentclass[aps,pra,twocolumn,groupedaddress,amsmath,amssymb,10pt]{revtex4-1}

\usepackage{epsfig}
\usepackage{amssymb}
\usepackage{amsmath}
\usepackage{graphicx}
\usepackage{amsfonts}
\usepackage{epsfig}
\usepackage{revsymb}
\usepackage{bm}
\usepackage{amsmath}
\usepackage{amssymb}
\usepackage{latexsym}
\usepackage{hyperref}
\usepackage{cleveref}

\begin{document}

\title{Quantum nucleation of gas bubbles in liquid helium }

\author{S.N. Burmistrov}
\author{L.B. Dubovskii}
\affiliation{NRC ``Kurchatov Institute'', 123182 Moscow, Russia}


\begin{abstract}
Liquid helium under negative pressures represents a unique possibility for studying 
nucleation and growth dynamics of cavities at low temperatures down to absolute zero. We analyze the growth dynamics of cavities and determine the temperature behavior of quantum cavitation rate and the crossover temperature between the classical and quantum cavitation mechanisms. The energy dissipation processes, which differ in kind in normal fluid $^3$He and superfluid $^4$He, result in the different temperature behavior of cavitation rate in $^3$He and $^4$He below the thermal-quantum crossover temperature. 

\end{abstract}
\maketitle

\section{Tensile strength of the liquid and its relation with the bubble cavitation}
\par
The detailed study of cavitation in liquids is of scientific and practical interest since the application of cryogenic liquids grows every year. The cavitation phenomena can have  harmful and dangerous outcome, e.g. in liquid hydrogen energy facilities, as well as useful examples, e.g. in bubble cameras for registering ionized particles. Over the last two decades, exploration has intensively been implemented for cavitation in liquid helium at low temperatures. The attractiveness of liquid He for studying the cavitation phenomenon is not so much in the absence of impurities and gas nuclei but in ability to study cavitation down to zero temperature when any thermal fluctuations are frozen out.   
\par
Usually, the tensile strength of liquid is understood as 
\begin{equation}\label{f51}
\Delta P= P_{\text{SVP}}-P_0
\end{equation}
where $P_{\text{SVP}}$ is the saturated vapor pressure above the flat surface and  $P_0$ is the pressure at which the voids or discontinuities in homogeneity  appear. We consider the specific features and distinctions of cavitation bubble inception in superfluid $^4$He and normal fluid $^3$He at sufficiently low temperatures when the quantum fluctuations become essential. 
\par
The first studies and tensile strength calculations for $^4$He are performed at high temperatures $T\gtrsim$ 1.5~K within the framework of classical thermal activation theory of homogeneous nucleation. The observed growth of tensile strength to 8 bars with lowering the temperature can well be agreed with the thermal activation theory for the rate of inception of gas bubbles in the stretched liquid 
 \cite{Niss}.  According to the first estimates \cite{Akul}, the thermal activation mechanism of bubble inception in liquid $^4$He should be succeeded by the quantum mechanism associated with the underbarrier tunneling of nuclei at about 310~mK. The tensile strength should become temperature-independent and reach about 15~bar.
\par
Later \cite{Xio,Xion},  the important physical circumstance related to
the liquid-gas phase transition was pointed out. This is an existence of spinodal where the liquid becomes absolutely unstable with respect to the long-wave density fluctuations due to vanishing the sound velocity. The latter is supposed to vanish near the spinodal pressure $P_c$ as 
\begin{equation}\label{f52}
c(P)\propto (P-P_c)^{\nu}
\end{equation}
and, correspondingly, the density approaches the critical one as 
\begin{equation}\nonumber
\rho -\rho _c\propto (P-P_c)^{1-2\nu}.
\end{equation}
The extrapolation of sound velocity from the positive pressure region to the negative one 
\cite{Xio,Xion,Mar,Mari} gives critical exponent $\nu$  close to values between 1/3 and 1/4. The critical zero temperature pressure is estimated as $P_c = - $(8.5 -- 9.5) bar for $^4$He  and 
$P_c = - $3 bar in  $^3$He. According to \cite{Mari}, the thermal-quantum crossover temperature $T_q$ should be about 220~mK and 120~mK in $^4$He and $^3$He, respectively. The further numerical calculations, using the local density functional and Monte Carlo 
\cite{Cau,Soli,Gui, Jez,Bor,Guil}, have shown qualitative and approximate quantitative agreement with the first results. Note that all these calculations of the quantum cavitation rate are based only on the thermodynamic properties of a liquid under full neglect of its relaxation or dynamic properties. Below we consider the cavitation more carefully with taking the energy dissipation processes and finite compressibility into account  \cite{Bur,Burm}. 
\par
The bubble nucleation in $^4$He under negative pressures has been studied experimentally down to 65~mK in \cite{Lam}. The results can be consistent with the idea that the cavitation results from the quantum tunneling across a potential barrier at temperatures below 600~mK. Quantum nucleation occurs at pressure $-$9.23 bar. For normal  $^3$He, though the threshold observed agrees with the estimate $P_c=-$3.1~bar, the results are too ambiguous to assert the presence of quantum cavitation regime above 40~mK. 
\par
We underline two points that allow us to clarify a discrepancy between theory \cite{Mari} and experiment \cite{Lam}. First of all, there is an energy dissipation during the underbarrier evolution of gas-like bubble in normal fluid $^3$He, reducing the quantum nucleation rate. Second, the experiments are performed near the spinodal where the sound velocity is small and vanishes. In the latter case the kinetic energy of the growing bubble decreases and the crossover temperature $T_q$ increases \cite{Burmi,Kor}. This can be a reason why temperature $T_q$ equals 0.6~K in superfluid 
$^4$He \cite{Lam,Cau} instead of 200~mK as is predicted in \cite{Xio}.
\par 
Recently, the systematic studies get started on the cavitation and growth of single electron bubbles in liquid helium \cite{Yad,Yada}.  In this case due to repulsive potential of about 1~eV it is energetically favorable for an electron to emerge a cavity free from helium atoms within the radius of about 19\,\AA\, \cite{Xing}. First of all, the injection of such single electron bubbles allows one to reduce significantly the cavitation threshold \cite{Gho}. The point is that the electron bubbles play a role of prepared nucleation centers for inception of cavitation gas bubbles and thus the cavitation acquires the specific features inherent in the heterogeneous nucleation.  

\section{Growth dynamics of bubbles in the liquid}

\par
The growth of a bubble in liquid as well as its nucleation is a complicated process. The bubble growth 
occurs in a condensed medium and is accompanied with both  dissipationless and energy dissipation processes involving the inhomogeneous outflow of liquid from the bubble, viscosity, heat conduction, and sound emission. The latter is due to finite compressibility of a liquid.  Even for the spherical bubble expanding uniformly in all directions, the derivation of general growth equation valid for an arbitrary growth rate is a complicated problem. Thus we make a series of simplifying assumptions.  
\par
We consider normal liquid,  say $^3$He, at an arbitrary pressure $P$.  Then we suppose the bubble to have a spherical shape of radius $R(t)$ and its radius grows at rate $\dot{R}(t)$. We neglect a possible existence of vapor inside the bubble since the vapor density in helium at low temperatures is negligibly small as compared with that of liquid phase. Then we can treat the bubble in the thin interfacial layer 
approximation. In other words, we describe the liquid-vacuum interface in terms of surface tension  
$\sigma$. This is justified if the bubble radius is much larger than the interface thickness. 
 \par
 The total energy of the system reads 
 \begin{equation}\label{f53}
 \mathcal{E}=\int _{r>R}d^3r\biggl[\frac{1}{2}\rho (\bm{r})\bm{v}^2(\bm{r})+\rho(\bm{r})\epsilon (\rho (\bm{r}))\biggr]+4\pi\sigma R^2
 \end{equation}
where $\bm{v}(\bm{r})$ and $\rho (\bm{r})$ are the velocity and liquid density at point $\bm{r}$. The first term is a sum of kinetic and internal energies,  $\epsilon$ being the internal energy per unit mass. The second term is the surface energy. To simplify the problem further, we neglect all heat effects that could accompany the bubble growth. 
\par
Deriving the growth equation, we use the continuity of mass flow across the interface. This yields at the interface $r=R(t)$ 
\begin{equation}\label{f54}
v(R)=\dot{R}(t).
\end{equation}
Then we employ the continuity of momentum flux density at the interface 
\begin{equation}\label{f55}
P(R)+\tau _{rr}(R)+2\sigma /R =0. 
\end{equation}
Here $P(R)$ is the pressure and $\tau _{rr}(R)$ is the radial components of viscous stress tensor at the interface. The viscous tensor $\tau _{ik}$ is given by 
\begin{equation}\label{f55a}
\tau _{ik}=-\,\eta\left(\frac{\partial v_i}{\partial x_k}+\frac{\partial v_k}{\partial
x_i}-\,\frac{2}{3}\delta _{ik}\frac{\partial v_l}{\partial x_l}\right)-\zeta\delta _{ik}
\frac{\partial v_l}{\partial x_l}
\end{equation}
where $\eta$ and $\zeta$ are the coefficients of viscosity. The last term in  (\ref{f55}) takes the Laplace pressure into account. 
\par
The boundary condition (\ref{f55}) in essence is the growth equation. We should express the pressure and viscous tensor at the boundary in the variables describing the bubble growth, i.e. $R(t)$ and $\dot{R}(t)$.  For this purpose, we use the continuity equation 
\begin{equation}\label{f56}
\frac{\partial\rho}{\partial t}+\text{div}\,\rho\bm{v} =0
\end{equation}
and the Navier-Stokes equation 
\begin{equation}\label{f57}
\rho\left(\frac{\partial\bm{v}}{\partial t}+(\bm{v}\!\cdot\!\nabla)\bm{v}\right)=-\,\nabla P+\eta\nabla
^2\bm{v}+\left(\zeta +\frac{\eta}{3}\right)\nabla(\text{div}\,\bm{v}) .
\end{equation}
The derivation of the general analytic solution  (\ref{f55}) -- (\ref{f57}) for an arbitrary behavior of growth rate $\dot{R}(t)$ in time $t$ is not a simple problem. We restrict ourselves with the limit of sufficiently small growth rates $\dot{R}/c\ll 1$, $c$ being the sound velocity. We maintain below the time derivatives in $R(t)$ to third order. Each term of expansion will have own physical meaning and its contribution depends on several factors as radius, growth rate, temperature, and kinetic coefficients.   
\par
To solve Eqs. (\ref{f56}) and (\ref{f57}), it is convenient to introduce the velocity potential 
$\bm{v}=\nabla\varphi$. Then, in our approximation the motion of liquid reduces to the linear equation corresponding to the propagation of sound with damping 
\begin{equation}\nonumber
\nabla ^2\varphi-\,\frac{\ddot{\varphi}}{c^2}+\,\frac{(4/3)\eta +\zeta}{\rho c^2}\nabla ^2\varphi =0 .
\end{equation}
The general solution reads for the sound propagating from the bubble and vanishing at the infinity
\begin{gather*}
\varphi (r,t)=\int _{-\infty}^{\infty}\frac{d\omega}{2\pi}\, e^{-i\omega t}\,\frac{e^{-i\lambda
_{\omega}r}}{r}\,\varPhi (\omega), 
\\
\lambda _{\omega}=\frac{\omega}{c}\left(1-2i\gamma _{\omega}\frac{c}{\omega}\right) ^{-1/2}\! , 
\;\;\;\gamma _{\omega}=\frac{4\eta /3 +\zeta}{2\rho c^3}\omega ^2.
\end{gather*}
Here $\gamma _{\omega}$ is the sound absorption coefficient due to viscosity. Unknown function 
$\varPhi (t)$ should be determined from the boundary condition (\ref{f54}). Within our approximation it is sufficient to take into account the first time derivative alone 
\begin{equation}\nonumber
\varPhi (t)=-\frac{\dot{V}(t)}{4\pi}, \;\;\; V(t)=\frac{4\pi}{3}R^3(t).
\end{equation}
Here $V(t)$ is the volume of an expanding bubble. Using the expression for the pressure in liquid
\begin{equation}\nonumber
P(r,t)-P=-\rho\dot{\varphi}-\,\rho\frac{(\nabla\varphi )^2}{2}+\biggl(\frac{4}{3}\eta +\zeta\biggr)\nabla
^2\varphi
\end{equation}
and Eq.  (\ref{f55a}), we find the growth equation   
\begin{equation}\label{f58}
\frac{2\sigma}{R}+P+4\eta\frac{\dot{R}}{R}+\rho\biggl(R\ddot{R}+\frac{3}{2}\dot{R}^2\biggr)-\frac{\rho}{4\pi c}\dddot{V}(R) +\ldots =0, 
\end{equation}
$P$ being the external pressure. In the lack of the surface, viscous and sound terms the bubble growth equation was first derived by Lord Rayleigh and later generalized by Plesset for the surface tension contribution. 
\par
Next, multiplying Eq.~(\ref{f58}) by $4\pi R^2$, we can represent the bubble growth equation in the general form describing the slow growth of nuclei in an arbitrary condensed medium as an expansion in growth rate $\dot{R}$
\begin{gather}
U'(R) + \mu _1(R)\dot{R} + \mu _2(R)\ddot{R} +\frac{1}{2}\mu _2^{\prime}(R)\dot{R}^2
 - \mu_3(R)\dddot{R} \nonumber 
 \\
- \frac{3}{2}\mu _3^{\prime}(R)\ddot{R}\dot{R}-\frac{1}{2}\biggl(\mu
_3^{\prime\prime}(R)-\frac{1}{2}\frac{\mu _3^{\prime 2}(R)}{\mu _3(R)}\biggr)\dot{R}^3 + \dots =0.   
\label{f59}
\end{gather}
The term $U'(R)$, finite at $\dot{R}=0$, has the meaning of the bubble energy  
\begin{equation}\label{f510}
U(R)=\frac{4\pi}{3}PR^3+4\pi\sigma R^2.  
\end{equation}
For the negative pressures, the bubbles of radius larger than 
$R_c=3\sigma /\vert P\vert $
prove to be energetically more favorable and the cavitation bubble growth becomes barrier-free. 
\par
The second term is the direct analog of the Stokes force with the friction coefficient equal to 
\begin{equation}\label{f512}
\mu _1(R)=16\pi\eta R .
\end{equation}
Note that equation  (\ref{f512}) refers to the hydrodynamic regime  implying  the smallness of mean free path $l(T)$ as compared with the bubble radius. On the contrary inequality $R\gg l(T)$,  the ballistic regime takes place and the friction coefficient is proportional to the bubble surface area. 
\par
The terms with the second derivative and the square of the first one are associated with the variable effective mass of a bubble 
\begin{equation}\label{f514}
\mu _2(R)=4\pi\rho R^3.
\end{equation}
These terms correspond to the kinetic energy of a bubble.  
\par
The rest terms of third order are originated from the finiteness of sound velocity or compressibility of a medium.  The corresponding term equals 
\begin{equation}\label{f515}
\mu _3(R)=\frac{4\pi\rho}{c} R^4.
\end{equation}
Obviously, the lower the sound velocity, the more the effect of this term on the cavitation dynamics. 
\par
One can see another physical aspect of nucleus growth, representing the equation as a derivative of energy 
\begin{gather*}
\frac{d}{dt}\biggl( U(R)+\frac{1}{2}\mu _2(R)\dot{R}^2 -\mu _3(R)\dot{R}\ddot{R}-\frac{1}{2}\mu
_3'(R)\dot{R}^3\biggr) 
\\
= -\mu _1(R)\dot{R}^2 -\,\frac{\rho}{4\pi c}\ddot{V}^2.
\end{gather*}
Hence, the right-hand side of equation is an energy dissipation function. The first term corresponds to the usual Ohmic dissipation with the variable friction coefficient. The second term results from changing the nucleus volume $V$ and  equals  the total power of sound emitted with a body immersed into liquid provided that the wavelength $\lambda$ of sound emitted is much larger than the size of a body, i.e. $\lambda\gg R$. In our case the latter is equivalent to inequality $\dot{R}\ll c$. 

\section{Energy dissipation and sound emission at the quantum cavitation }
\par
Here we estimate the temperature of thermal-quantum crossover and the nucleation rate of bubbles at zero temperature. According to the general theory of quantum nucleation, the cavitation rate is given by the expression 
$$
\Gamma =\Gamma _0\exp (-S)
$$
where $S$  is the extremum of effective action $S_{\text{\text{ef}}}[R_{\tau}]$ corresponding unambiguously to the bubble growth equation. The general expression for the effective action reads
\begin{multline*}
S_{\text{ef}}[R(\tau )] 
= \!\!\!\int\limits _{-1 /2T}^{1 /2T}\!\!\! d\tau\,
\biggl\{\frac{\mu_2(R_{\tau})\dot{R}_{\tau}^2}{2}+U(R_{\tau})
\\
+\frac{1}{4\pi}\int\limits _{-1 /2T}^{1/2T}\!\!\! d\tau '\,
\left[\gamma_1(R_{\tau })-\gamma _1(R_{\tau ')}\right]^2\frac{(\pi T)^2}{\sin ^2\pi T(\tau -\tau ')}
\\
-\frac{1}{4\pi}\int\limits _{-1 /2T}^{1 /2T}\!\!\! d\tau '\, \biggl[\frac{\partial
\gamma_3(R_{\tau })}{\partial\tau}-\frac{\partial\gamma _3(R_{\tau '})}{\partial\tau '}\biggr]^2\frac{(\pi T)^2}{\sin ^2\pi T(\tau -\tau ')}\biggr\}.
\end{multline*}
The relation of vertices $\gamma _1(R)$ and $\gamma_3(R)$ to the kinetic coefficients  $\mu_1(R)$ and $\mu_3(R)$ is given by 
$$
\bigl(\partial\gamma_1(R)/\partial R\bigr)^2=\mu_1(R)\;\;\;\text{and}\;\;\;\bigl(\partial\gamma_3(R)/\partial R\bigr)^2=\mu_3(R).
$$
\par
Below we focus on the ballistic regime when the mean free path $l(T)$ exceeds significantly the quantum critical bubble radius $R_c$. This regime is more appropriate to the low temperature experimental situation. In fact, at low temperatures $T<1$ K the mean free path increases significantly in normal $^3$He and superfluid $^4$He. The opposite hydrodynamical regime $R_c>l(T)$ requires the larger magnitudes of critical bubble radius. The large critical radius $R_c$ results in such negligible nucleation rate that the homogeneous cavitation becomes unobservable on the reasonable experimental time scales. The lack of hydrodynamical regime implies inequality $T_l(R_c)>T_q(R_c)$ where 
$T_l$ is the temperature when the mean free path equals $R_c$ and  $T_q$ is the thermal-quantum crossover temperature. 
\par
In the ballistic $R_c<l(T)$ regime the effective action is convenient to rewrite as 
\begin{multline}\label{f516}
S_{\text{ef}}[R_{\tau }]
 =  \int\limits_{-1 /2T}^{1/2T}\!\!\! d\tau\,
\biggl(\frac{4\pi}{3}PR^3_{\tau}+4\pi\sigma R^2_{\tau}+2\pi\rho R^3_{\tau}\dot{R}^2_{\tau}\biggr)
\\ 
+\frac{1}{4\pi}\iint\limits _{-1/2T}^{\, 1/2T}\!\!\!d\tau d\tau '   \biggl\{ \frac{\rho}{4\pi}u\bigl[
A(R_{\tau})-A(R_{\tau '})\bigr] ^2 
\\
-\frac{\rho}{4\pi c}\bigl[\dot{V} (R_{\tau})-\dot{V}(R_{\tau '})\bigr] ^2\biggr\}\frac{(\pi T)^2}{\sin ^2\pi T(\tau -\tau ')}.
\end{multline}
Here $A(R)=4\pi R^2$ is the bubble surface area,  $V(R)=4\pi R^3/3$  is the bubble volume and  $u=\alpha\eta /\rho l$ equals approximately the typical velocity of elementary excitations in the liquid. The value of numerical coefficient $\alpha$ is about unity. For normal $^3$He, the magnitude for $u$ is about the Fermi velocity  and the possible temperature corrections to zero temperature case are about  $(T/T_F)^2$ where $T_F$ is the degenerate temperature. 
\par
In superfluid $^4$He, where the energy dissipation is associated with the presence of normal component alone, the behavior of parameter $u$ is different 
\begin{equation}\label{f517}
u(T)\approx c \rho _n(T)/\rho .
\end{equation}
Here $\rho _n(T)$ is the normal component density and is determined with phonons at temperatures  $T<$0.5 K
\begin{equation}\label{f517a}
\rho _n(T)=\frac{2\pi ^2}{45}\,\frac{T^4}{c^5}.
\end{equation}
\par
The high temperature region yields the stationary path $R_{\tau}\equiv R_0=2R_c/3$ and classical action $S_{\text{cl}}=U_0/T$ with the Arrhenius law for the cavitation rate 
\begin{equation}\label{f518}
\Gamma =\Gamma _0\exp (-U_0/T)\; , \;\;\; U_0=\frac{16\pi\sigma ^2}{3\vert P\vert^2}. 
\end{equation}
\par
We start the study of low temperature quantum behavior for the cavitation rate from analyzing the stability of classical path. Let us represent an arbitrary path as 
\begin{equation}\nonumber
R(\tau )=R_0+r(\tau)
\end{equation}
and expand $r(\tau )$ into a Fourier series over frequencies 
\begin{equation}\nonumber
r(\tau )=T\sum \limits_{n}r_n e^{-i\omega\tau} ,\;\; \omega _n =2\pi T ,\;\; n=0,\pm 1, \pm 2,\ldots
\end{equation}
We have for small $r_n$
\begin{equation}\nonumber
S_{\text{ef}}=\frac{U_0}{T}+\frac{1}{2}T\sum \limits_{n}\alpha _n\vert r_n\vert^2+\ldots
\end{equation}
The coefficients of expansion are given by 
$$
\alpha _n=U''_{0}+16\pi\rho uR_0^2\vert\omega _n\vert+4\pi\rho R_0^3\omega _n^2-\,\frac{4\pi\rho R_0^4}{c}\vert\omega _n\vert ^3. 
$$
As the temperature lowers, coefficients  $\alpha _{\pm 1}$ vanish at temperature $T_1$ determined by 
\begin{equation}\label{f519}
-\sigma +4\rho uR_0^2\omega _1 +\rho R_0^3\omega _1^2 -\frac{\rho R_0^4}{c}\omega _1^3=0\; , \;\;
T_1=\frac{\omega _1}{2\pi}.
\end{equation}
Below temperature $T_1$ the classical path becomes absolutely unstable with respect to the oscillations of mode $r_{\pm 1}$. 
\par
 The thermal-quantum crossover temperature $T_q$  equals $T_1$ if the effective action  goes over smoothly to the Arrhenius exponent or becomes somewhat higher than $T_1$ if the transition from the classical path to quantum one has a jump-like character, i.e.   $T_q\geqslant T_1$. For our purposes, the approximate estimate $T_q\approx T_1$ is quite sufficient. Using  (\ref{f519}) in the large radius limit $R_0\rightarrow\infty$ or small negative pressures $\vert P\vert\rightarrow 0$, we find the following estimate: 
\begin{equation}\label{f520}
T_q\approx \frac{\hbar\sigma}{8\pi\rho uR_0^2}=\frac{\hbar}{32\pi\sigma\rho u}\vert P\vert^2.
\end{equation}
To be fully correct, we imply that the bubble growth rate should be smaller as compared with the velocities of excitations and sound. Since the typical time of underbarrier growth is $(2\pi\omega _1)^{-1}$, we arrive at 
\begin{equation}\label{f521}
\frac{2\pi\omega _1R_0}{u}=\frac{\pi\sigma}{2\rho u^2R_0}\ll 1.
\end{equation}
This inequality restricts the pressure magnitude $\vert P\vert\ll \rho u^2$
when our approximation is satisfied. If Eq.  (\ref{f521}) breaks down, the terms of higher order than 
$\omega_1$   are to be introduced in (\ref{f519}). 
\par
Unlike normal $^3$He, in superfluid $^4$He the normal component density $\rho _n(T)$ vanishes as $T\rightarrow 0$  and the contribution of Ohmic term in (\ref{f519}) decreases. Rewriting  (\ref{f519})  
with the aid of (\ref{f517a}), we have 
\begin{equation}\label{f523}
-\sigma +\frac{R_0^2}{90\pi ^2c^4} +\rho R_0^3\omega _1^2 -\,\frac{\rho R_0^4}{c}\omega _1^3=0.
\end{equation}
The condition $\omega _1R_0/c\ll 1$ is supposed to be satisfied. 

As is seen, the dissipative Ohmic term has no essential effect on the crossover temperature under  $R_0\gg R_{*}$ where
\begin{equation}\label{f524}
R_{*}\approx\left(\frac{\hbar ^2\sigma ^3}{8100\pi ^4 c^8\rho ^5}\right) ^{1/11}.
\end{equation}
For radii $R_0\gg R_{*}$, the crossover temperature proves to be approximately the same as it follows from the dissipationless cavitation model \cite{Mari,Akul}
\begin{equation}\label{f525}
T_q\approx \frac{\hbar}{2\pi}\sqrt{\frac{\sigma}{\rho
R_0^3}}=\frac{\hbar}{4\pi}\,\frac{\vert P\vert^{3/2}}{\sqrt{\alpha\rho}}.
\end{equation}
To satisfy the approximation of low growth rates $\omega _1R_0\ll c$, we should put limitations  for 
$R_0$ or pressure $P$
\begin{equation}\label{f526}
R_0\gg \sigma /(\rho c^2) \;\; \text{or}\;\; \vert P\vert\ll\rho c^2.
\end{equation}
Numerically, if the $^4$He  parameters are taken at zero pressure, we find  $R_{*}=\sigma /\rho c^2\approx$ $0.5$~\AA.
Thus, estimate (\ref{f525}) is associated with the applicability of macroscopic description  implying the bubble radius larger than the interface thickness. Note that condition  (\ref{f526}) will not be satisfied in the vicinity of spinodal.

Let us turn to the low temperature region $T\ll T_q$ and consider first the case of normal $^3$He, assuming the velocity $u(T)=\text{const}$. Since we retain the approximation of low growth rates, the main contribution to the effective action is due to energy dissipative term nonlocal in time.  The other two dynamical terms can be treated as a perturbation. For zero temperature, we have approximately
\begin{equation}\label{f527}
S(T=0)\approx 4\pi\rho u R_c^4\biggl( 1+\frac{\sigma}{2\rho u^2R_c}-\frac{u}{9c}\biggl(\frac{\sigma}{\rho
u^2R_c}\biggr) ^{\! 2}\biggr)
\end{equation}
and $S(T=0) \propto \vert P\vert ^{-4}$. 
This answer is a decomposition in $\dot{R}/u\ll 1$ if we take into account that the typical growth time of bubble is about $\rho uR_c^2/\sigma$.

Unlike the dissipationless kinetics \cite{Mari,Akul}, the energy dissipation of a bubble in the course of its growth results in the effective action  with the kinetic terms depending explicitly on the temperature.  Thus, the cavitation rate can be temperature-dependent and we can expect  \cite{Burmis} that 
\begin{equation}\label{f528}
\Delta S(T)=S(T)-S(0)\approx -S(0)(T/T_q)^2.
\end{equation}
The temperature corrections affect noticeably the cavitation rate while $\vert\Delta S(T)\vert >\hbar$, i.e. at temperatures $T>T_2$
\begin{equation}\label{f529}
T_2\sim \frac{\sigma}{R_c^4}\left(\frac{\hbar}{4\pi\rho u}\right) ^2\propto \vert P\vert ^4.
\end{equation}
Therefore, in region $T_2<T\ll T_q$ the cavitation rate should follow $\log [\Gamma
(T)/\Gamma (0)]\sim T^2$.

Unlike normal $^3$He,  the normal component density in superfluid $^4$He vanishes rapidly with approaching zero temperature and the quantum cavitation is governed mainly by the well-known dissipationless term associated with the kinetic energy of a bubble in the Rayleigh-Plesset Lagrangian  \cite{Mari,Akul}. Involving the finite sound velocity correction, we find  for $T=0$
\begin{equation}\label{f530}
S(T=0)=\frac{5\sqrt{2}\pi ^2}{16}(\sigma\rho ) ^{1/2} R_c^{7/2}\biggl( 1-\frac{4}{9c}\sqrt{\frac{2\sigma}{\rho R_c}}\biggr).
\end{equation}
The order of the magnitude for the second term is a ratio of the underbarrier growth rate to the sound velocity.

The temperature dependence of action $S(T)$ is associated with the normal component density and sound term. The both terms yield the same temperature behavior, however, the contributions are of opposite signs   
\begin{equation}\label{f531}
\Delta S(T)=S(T)-S(0)=\biggl(\frac{8\pi ^3\hbar}{45c^3}-\frac{\rho
^2R_c^5}{\sigma}\biggr)\frac{R_c^4}{c}\biggl( \frac{T}{\hbar}\biggr) ^4. 
\end{equation}
In the pressure region $\vert P\vert<3\sigma /R_{*}$ or for the quantum critical radii satisfying 
\begin{equation}\label{f532}
R_c>R_{*}=\left(\frac{8\pi ^3}{45}\,\frac{\sigma\hbar}{\rho ^2 c^3}\right) ^{1/3}, 
\end{equation}
the sound term dominates over. For the estimate, let us take the $^4$He parameters at zero pressure. The numerical estimate gives $R_{*}\approx$ $2.4$ \AA\, comparable with interatomic spacing $a$. Thus, in the region of macroscopic description $R_c\gg a$ the normal component contribution is small and, therefore, the cavitation rate $\Gamma (T)$ should enhance as the temperature increases. Let us estimate the temperature when the temperature dependence becomes essential, i.e. $\vert\Delta S(T_2)\vert\approx\hbar$. From (\ref{f531}) we have 
\begin{equation}\label{f533}
T_2\sim \hbar \left(\frac{\hbar\sigma c}{\rho ^2R_c^9}\right)^{1/4}.
\end{equation}
However, temperature $T_2$ is lower than the thermal-quantum one $T_q$ only for the critical radii exceeding 
\begin{equation}\label{f534}
R_2=\frac{4\pi}{3}\left(\frac{16\pi\hbar c}{3\sigma}\right) ^{1/3}.
\end{equation}
The estimate   gives $R_2\approx$ $40$ \AA\, for zero pressure parameters.  Thus only the large bubbles of radius $R_c>R_2$ have the temperature range $T_2<T<T_q$, in which $\vert\Delta S(T)\vert>\hbar$. For the critical bubbles $R_c<R_2$, the scale of varying $\log \Gamma (T)/\Gamma (0)\propto T^4$ is insignificant. 

\section{Sound emission at quantum cavitation in superfluid $^4$H{\scriptsize e}}

The cavitation in liquid $^4$He is consistently of interest \cite{Lam,Cau}. The attractiveness  of study is associated with the lack of impurities which usually play a role of heterogeneous nucleation centers. The cavitation process is induced with the sound 1 MHz pulses focused at the center of experimental cell. The sound pulses produce the pressure oscillations of several bars in amplitude. 
The size of experimental cell is about 8 mm and approximate size of sound focus is 0.12 mm. Cavitation occurs stochastically. At the fixed temperature and ambient pressure some sound pulses of the given amplitude results in nucleating the cavitation bubbles but the others no. Applying a series of pulses
and fixing the number of cavitation events, one can determine the cavitation probability as a function of external parameters, namely, applied voltage, temperature, and external pressure.  According to 
\cite{Lam,Cau}, the probability or cavitation rate is clearly temperature-dependent above approximately 400 mK. One of the problems in interpreting the experiments is associated with the following. The point is that there is a sound absorption maximum in this temperature region. On account of this fact the cavitation rate becomes temperature-independent up to 600 mK. Above this temperature the cavitation threshold decreases as the temperature grows, corresponding to the thermal activation mechanism. This result is interpreted as a quantum-thermal crossover in cavitation. Note that the stochasticity at the temperature-independent threshold of cavitation does not correspond to the assumption about achieving the spinodal pressure. 

One more complication in interpreting the experiments is associated with following. The temperature measured in the cell can differ from that in the sound wave focus where the cavitation bubble nucleates. 
The point is that the acoustic wave is adiabatic in first approximation and, therefore, the temperature in the focus oscillates together with the pressure. On the assumption of adiabaticity the temperature can be
estimated \cite{Lam}. For the temperature lower 600 mK, the phonons contribute mainly to the entropy 
\begin{equation}\nonumber
S\approx S_{\text{ph}}=\frac{2\pi ^2T^3}{45\rho ^3c^3}.
\end{equation}
In the isoentropic process the temperature is proportional to sound velocity $c$.  As is shown  in \cite{Lam}, in the vicinity of spinodal at 
 $P=-$ 9.23~bar the sound velocity is about 74~m/s. This is three times as smaller as compared with zero pressure.  Hence the local temperature in the focus may be smaller by the same three times than the temperature in the surrounding cell \cite{Lam}.

This interpretation is self-consistent provided the two following conditions are fulfilled. The first is connected with the well-known fact that the nonlinear effects originate at very early stage \cite{Ivl}.  The typical field when nonlinear effects appear is proportional to 
\begin{equation}\nonumber
V=\tilde{V}\exp (\omega t_0). 
\end{equation}
Here $\omega$ is the sound frequency, $t_0$ is  the time of underbarrier evolution, and $\tilde{V}$ is the sound amplitude.  In experiment \cite{Lam} $\omega =$ 1~MHz and $t_0$ can be estimated as  $t_0\sim 10^{-10}$~s in the experimentally analyzed vicinity to the spinodal. Thus, we expect $\omega t_0\ll 1$  in experiment and the nonlinear effects can be neglected. The second condition is associated with the time of setting the system into equilibrium and can be represented as
\begin{equation}\nonumber
\omega\tau\ll 1 \;\;\;\; \text{or} \;\;\;\; l\ll\lambda 
\end{equation}
where $\tau$ is the collision time between excitations in $^4$He. This means also the smallness of mean free path $l$ compared with the sound wavelength $\lambda$. In experiment \cite{Lam} the inverse limit takes place for the whole temperature range. According to \cite{Mor}, the phonon mean free path will exceed the acoustic focus size ($\sim$ 0.12~mm) at about 0.8~K and for $T<0.5$~K  becomes more than the cell size 8~mm. As a result, the local temperature in the sound wave cannot follow the pressure oscillations in the temperature region considered and the condition of entropy constancy breaks down. So, the local temperature in the wave focus equals the temperature beyond the focus.  Correspondingly, the thermal-quantum crossover temperature should equal 0.6~K.

Let us compare the experimental crossover temperature with the calculated one. The simple estimate, based on the dissipationless model for the thin wall bubble in $^4$He, gives $T_q=$ 0.15~K. Here we use  $\rho =$0.095~g/cm$^3$,  $\sigma = $0.37~erg/cm$^2$, and experimental magnitude $P = -$ 9.5~bar. The calculations using the density functional theory \cite{Mari} give $T_q=$ 0.2~K. The insignificant difference between these two approximations is not surprising since the both are based on the same magnitude of surface tension. Moreover, this difference is smaller than the experimental magnitude which should be taken as 0.6~K instead of 0.2~K as in \cite{Lam}. These results are insensitive to the presence of energy dissipation term in the effective action.

As it concerns the sound term reducing the effective action by a factor  (\ref{f530})
$$
1-(4/9c)\sqrt{2\sigma /\rho R_c} = 1-\dot{R}/c,
$$
it, in general, is not small. The estimate gives 
$\dot{R}/c = $ 0.48. Correspondingly,  crossover temperature $T_q$ becomes twice as high  and approaches $T_q=$ 400~mK. So, we can assert that the tendency to increasing the crossover temperature due to finite compressibility is a reason for the higher magnitude  $T_q$ observed experimentally. 

\section{The effect of energy dissipation and viscosity at cavitation in\;\; normal $^3$H\scriptsize{e}}

Let us consider cavitation experiments in $^3$He \cite{Cau,Caup} representing normal viscid  fluid in the experimental temperature range 50 -- 1000 mK.  The simple estimate in the dissipationless model (\ref{f525}) gives $T_q=$ 90~mK for $\rho =$ 0.054~g/cm$^3$, $\sigma =$ 0.16~erg/cm$^2$, and pressure $P= -$ 3.1~bar near the spinodal one. More accurate calculation within the density functional theory \cite{Mari} yields somewhat higher crossover  temperature $T_q=$ 125~mK. However the crossover to quantum regime has not been observed down to 50~mK \cite{Cau,Caup}.

These both estimates ignore the fact that $^3$He is viscid fluid with viscosity $\eta \sim \rho
v_F^2\tau$ where $\rho$ is the density, $v_F$ is the Fermi velocity. The collision time is 
$\tau\sim D\hbar\epsilon _F/T^2$ where $\epsilon _F$ is the Fermi energy and $D\sim (p_Fa_0/\hbar)^{-2}$ is the dimensionless parameter expressed in terms of Fermi momentum 
$p_F$ and scattering length $a_0$. Thus the mean free path is  $l=A/T^2$. From \cite{Mor} we find that the mean free path is about $l\sim 40$~\AA\, at 
$T=0.15$~K and zero pressure.  Let us introduce temperature $T_l=\sqrt{A/R_c}$ separating the hydrodynamic regime of bubble growth from the ballistic one. Putting the typical quantum critical radius $R_c$ to be about 10 -- 20~\AA\, for the cavitation rates in experiment, we arrive at $T_l= 0.2 - 0.3$~K.  
Underline that temperature $T_l$  is above the quantum crossover temperature $T_q$ equal to about  0.1~K according to the estimates in the energy dissipationless models of cavitation.  Involving the energy dissipation will only decrease the crossover temperature $T_q$. In any case the cavitation in the quantum region should occur under ballistic bubble growth.

Let us compare contribution $S_{\text{kin}}$ from the dissipationless terms in the complete effective action (\ref{f527}), given by Eq. (\ref{f530}), with contribution $S_{\text{diss}}$ from the dissipative Ohmic term equal approximately to $S_{\text{diss}}\sim\rho uR_c^4\propto \vert P\vert^{-4}$. We find 
$$
\frac{S_{\text{kin}}}{S_{\text{diss}}}=\frac{5\sqrt{2}\pi ^2}{16\sqrt{3}}\sqrt{\frac{\vert P\vert}{\rho
u^2}}\biggl(1-\frac{4}{9c}\sqrt{\frac{2\vert P\vert}{3\rho}}\biggr) .
$$
This estimate means that the quantum cavitation in the pressure region $\vert P\vert \ll\rho u^2$ is accompanied with the strong energy dissipation and takes place in the viscous damped regime. The magnitude $u$ equals approximately the Fermi velocity, i.e. $\rho u^2$ is about 3 bar representing the scale of  spinodal pressure.  So, in $^3$He  it is necessary to involve energy dissipation in the whole region of possible pressures.  For $\vert P\vert\ll \rho u^2$, the thermal-quantum crossover temperature is given by 
$$
T_q\sim\frac{16\pi\hbar}{243}\,\frac{P^2}{\rho u\sigma}\propto P^2 .
$$
The point here is that the thermal-quantum crossover temperature $T_q$ proves to be much smaller than that in the energy dissipationless model. In this sense no observation of crossover in normal fluid $^3$He is amazing. For the negative pressures $P\approx -3$~bar close to the spinodal one, the viscous term becomes of the order of dissipationless one. The thermal-quantum crossover temperature should be a half of that obtained in the dissipationless models. As a result, the thermal-quantum crossover temperature in normal fluid $^3$He should not be higher than 50 mK.

Unlike the cavitation in superfluid $^4$He, the energy dissipation effects in normal fluid $^3$He play an essential role in the quantum cavitation since the normal excitation density does not vanish in the low temperature limit. The viscous dissipative effects lead to reducing both the quantum cavitation rate and the thermal-quantum crossover temperature as compared with the models neglecting the energy dissipation processes.  This conclusion agrees completely with the experimental studies of cavitation in normal fluid $^3$He.

In short, it is well observed \cite{Abe} that the growth dynamics of sound-induced bubbles in liquid helium is strongly dependent on whether the liquid is normal $^3$He, superfluid $^4$He, or liquid $^3$He-$^4$He mixture. In this paper we have attempted to motivate, define and discuss the question: what is the effect of dissipative processes in  liquid He on the quantum tunneling cavitation rate of bubbles at sufficiently low temperatures?


\begin{thebibliography}{99}

\bibitem{Niss} J. A. Nissen, E. Bodegom, L. C. Brodie, and J. S. Semura,   
Phys. Rev. B \textbf{40}, 6617 (1989).
\bibitem{Akul} V. A. Akulichev and V. A. Bulanov, Sov. Phys. Acoustics \textbf{20}, 501 (1975).
\bibitem{Xio} H. J. Maris and Q. Xiong, Phys. Rev. Lett. \textbf{63}, 1078 (1989).
\bibitem{Xion} Q. Xiong and H. J. Maris,   J. Low Temp. Phys. \textbf{77}, 347 (1989).
\bibitem{Mar} H. J. Maris, Phys. Rev. Lett. \textbf{66}, 45 (1991).
\bibitem{Mari} H. J. Maris,  J. Low Temp. Phys. {\bf 98},  403 (1995).
\bibitem{Cau} F. Caupin and S. Balibar, Phys. Rev. B \textbf{64}, 064507 (2001).
\bibitem{Soli} M. A.~Solis and J.~Navarro, Phys. Rev. B \textbf{45}, 13080 (1992).
\bibitem{Gui} M. Guilleumas, M. Pi, M. Barranco, J. Navarro, and M. A. Solis, Phys. Rev. B 
 \textbf{47}, 9116 (1993).
\bibitem{Jez} D. M. Jezek, M. Guilleumas, M. Pi, M. Barranco, and J. Navarro, Phys. Rev. B \textbf{48}, 16582 (1993).
\bibitem{Bor} J. Boronat, J. Casulleras, and J. Navarro, Phys. Rev. B,  \textbf{50}, 3427 (1994).
\bibitem{Guil} M. Guilleumas, M. Barranco, D. M. Jezek, R. J. Lombard, and M. Pi, Phys. Rev. B, \textbf{54}, 16135(1996).
\bibitem{Bur} S. N. Burmistrov and L. B. Dubovskii, Low Temp. Phys. \textbf{23}, 389 (1997).
\bibitem{Burm} S. N. Burmistrov and L. B. Dubovskii,  Zh. Exsp. Teor. Fiz.  \textbf{118}, 885 (2000).
\bibitem{Lam} H. Lambare, P. Roche, S. Balibar, H. J. Maris, O. A. Andreeva, C. Guthman,   K. O. Keshishev, and E. Rolley, Eur. Phys. J. B \textbf{2}, 381 (1998).
\bibitem{Caup} F. Caupin, P. Roche, S. Marchand, and S. Balibar, J. Low Temp. Phys. \textbf{113}, 473 (1998).
\bibitem{Burmi} S. N. Burmistrov and L. B. Dubovskii, J. Low Temp. Phys. \textbf{96}, 131 (1994).
\bibitem{Kor}  S.E. Korshunov,  Sov. J. Low Temp. Phys. \textbf{14}, 316 (1988).
\bibitem{Yad} N. Yadav, V. Vadakkumbatt, and A. Ghosh, J. Low Temp. Phys. \textbf{201}, 97 (2020).
\bibitem{Yada} N. Yadav, V. Vadakkumbatt,  H. J. Maris,  and A. Ghosh, J. Low Temp. Phys. \textbf{187}, 618 (2017).
\bibitem{Xing} Y. Xing and H. J. Maris J. Low Temp. Phys. \textbf{201}, 634 (2020).
\bibitem{Gho} A. Ghosh and H. Maris, J. Low Temp. Phys. \textbf{134}, 251 (2004).
\bibitem{Burmis} S. N. Burmistrov and L. B. Dubovskii, Phys. Lett. A \textbf{127}, 79 (1988) \bibitem{Ivl} B. I. Ivlev and V. I. Mel'nikov, Phys. Rev. Lett. \textbf{55}, 1614 (1985); in: \textit{Modern Problems in Condensed Matter Sciences}, (North-Holland, Amsterdam, 1992), Vol.34.
\bibitem{Mor} M. Morishita, T. Kuroda, A. Sawada, and T. Satoh, J. Low Temp. Phys. \textbf{76}, 387 (1989).
\bibitem{Abe}
H. Abe, M. Morikawa, T. Ueda, R. Nomura, Y. Okuda, and S.N. Burmistrov, J. Fluid Mech. \textbf{619}, 261 (2009).
\end{thebibliography}
\end{document}